\documentclass{article}
\usepackage{amsmath}
\usepackage[dvips]{epsfig}
\title{Running coupling for Wilson bermions
\thanks{Talk given at the International Symposium on Lattice Field
Theory, June 28--July 3, 1999, in Pisa, Italy.}
}

\author{Juri Rolf and Ulli Wolff\\[5mm]
Humboldt-Universit\"at zu Berlin,
             Institut f\"ur Physik,\\
              Invalidenstrasse 110,
              10115 Berlin}

\begin{document}
\maketitle
\begin{abstract}
A non perturbative finite size scaling technique is used to study
a running coupling in lattice Yang-Mills theory coupled to a 
bosonic Wilson spinor field in the Schr\"odinger functional
scheme. This corresponds to two negative flavours. The scaling
behaviour in this case is compared to quenched results and to QCD with
two flavours. The continuum limit is confronted with renormalized
perturbation theory.
\end{abstract}
\section{Introduction and motivation}
One of the main goals of the ALPHA collaboration is the non
perturbative computation of the strong coupling $\alpha_S$ for energy
scales ranging from hadronic scales to perturbative high energy
scales. To this end a
finite volume scale dependent renormalization scheme for
QCD has been
invented~\cite{Luscher:1992an,Luscher:1994gh}. Its
central object is the step scaling 
function $\Sigma$ which describes the running of the coupling under a
discrete change of the scale. Other ingredients include Schr\"odinger
functional boundary conditions, ${\rm O}(a)$ improvement and non
perturbative renormalization. This method has been used successfully
in the quenched approximation (see~\cite{Sommer:1997xw} for a review). First results for full QCD
with two flavours have been obtained recently~\cite{Jansen:1998mx}.
%
%
%
%
%
%
\begin{figure}[htb]
\vspace{-20mm}
\begin{center}
\epsfig{file=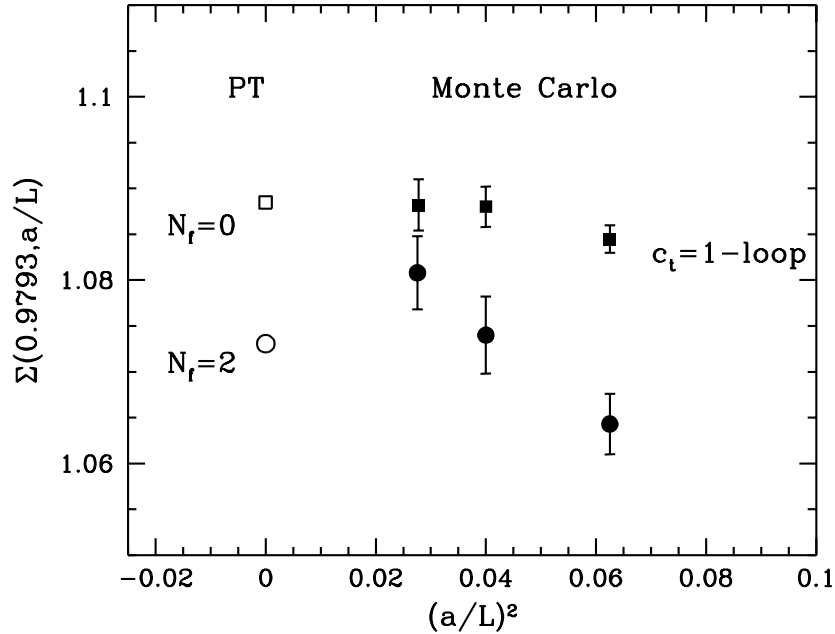,width=12.0cm}
\parbox{12.0cm}{
\vspace{-0.9cm}
\caption{\label{fig1a} \sl Preliminary results for the step scaling function in QCD
      with two flavours at $u=0.9793$ compared with the quenched
      approximation. Throughout this paper we denote $N_f=2$ by
      circles, $N_f=0$ by squares and $N_f=-2$ by triangles.}
}
\vspace{-1.0cm}
\end{center}
\end{figure}
In figure~\ref{fig1a} the results for the step scaling function at the
coupling $\bar{g}^2=0.9793$ are shown and compared with the
quenched approximation and the perturbative continuum
limit\footnote{These data were obtained in collaboration with A. Bode, R. Frezzotti,
      M. Guagnelli,  K. Jansen, M. Hasenbusch, J. Heitger,
      R. Sommer and P. Weisz.}. Here an
${\rm O}(a)$ improved action was used. The observed cut-off effects
are larger than in the quenched approximation. A naive extrapolation of the
Monte Carlo data to the continuum limit yields a value which lies
2-3\% above the perturbative estimate, although the coupling
$\bar{g}^2$ is expected to be small enough to be in the perturbative
regime. Furthermore, the cut-off effects in the quenched and in the
full theory computed in lattice perturbation theory to two loops are
of the same size~\cite{achimundulli}. These observations can be
interpreted as follows:
\begin{itemize}
\item statistical fluctuation
\item true 3\%\ deviation from renormalized perturbation theory
\item lattice artifacts too large for continuum extrapolation
\end{itemize}
The investigation of these possibilities in full QCD would consume a
substantial amount of computer time. Therefore we
use Yang-Mills theory coupled to a 
bosonic Wilson spinor field, which corresponds to setting $N_f=-2$
in the QCD partition function, as a toy model to 
study the extrapolation to the continuum limit for a system different
from pure gauge theory. This model has been called the bermion model
in the literature~\cite{Anthony:1982fe,deDivitiis:1995au}.
\section{The bermion model in the Schr\"odinger functional setup}
%
%
%
%
%
%
Let the space time be a hypercubic Euclidean lattice with lattice spacing
$a$ and volume $L^3\times T$. In the following we set $T=L$. The
$SU(3)$ gauge field $U(x,\mu)$ is defined on the links while the
bermion field $\phi(x)$ which is a bosonic spinor field with color and Dirac
indices is defined on the sites of the lattice. In the space
directions we impose periodic boundary conditions while in the time
direction we use Dirichlet boundary conditions. The
boundary gauge fields can be chosen such that a constant color
electric background field is enforced on the system which can be
varied by a dimensionless parameter $\eta$~\cite{Luscher:1992an}. 

As explained above our goal is to continue the exponent of the fermion
determinant in the QCD partition function to the negative value
$N_f=-2$. This can be achieved by integrating the
bosonic field $\phi$ with the Gaussian action
\begin{eqnarray}
 S_B&=&\sum_x \vert M\phi(x)\vert^2,\ \text{where}\\
 \frac{1}{2\kappa}M\phi(x)&=&({\cal D}+m_0)\phi(x).
\end{eqnarray}
$\cal D$ is the Wilson Dirac operator with hopping parameter
$\kappa=(8+2m_0)^{-1}$. For the gauge fields we employ the
action
\begin{equation}
  \label{eq:15}
   S_G=\frac{1}{g_0^2}\sum_p w(p) \text{tr}(1-U(p)).
\end{equation}
The weights $w(p)$ are defined to be one for plaquettes $p$ in the
interior and they equal $c_t$ for time like plaquettes attached to the
boundary. 
The choice $c_t=1$ corresponds to the standard Wilson action.
However, $c_t$ can be tuned in order to reduce lattice
artifacts. For the bermion case we have in this work always chosen
$c_t=1$. 

Now the Schr\"odinger functional is defined as the partition function
in the above setup:
\begin{eqnarray}
\lefteqn{
  Z=\int {\cal D}U{\cal D}\phi{\cal D}\phi^+\ e^{-S_G-S_B}}\\
   &=&\int {\cal D}U\ e^{-S_G(U)}\det(M^+M)^{N_f/2},
\end{eqnarray}
with $N_f=-2$. The effective action 
\begin{equation}
\Gamma=-\log Z
\end{equation}
with the perturbative expansion
\begin{equation}
\Gamma=g_0^{-2}\Gamma_0+\Gamma_1+g_0^2\Gamma_2+\ldots
\end{equation}
is renormalizable with no extra counterterms up to an additive divergent
constant. That means that the derivative  $\Gamma'=\frac{\partial
  \Gamma}{\partial\eta}$ is a renormalized quantity and
\begin{equation}
\bar{g}^2(L)=\Gamma_0'/\Gamma'\vert_{\eta=0}
\end{equation}
defines a renormalized coupling which depends only on $L$ and the
bermion mass $m$. Note that this coupling can be computed efficiently
as the expectation value $\frac{\partial \Gamma}{\partial
  \eta}=\left\langle\frac{\partial S}{\partial\eta}\right\rangle$. 
The mass $m$ is defined via the PCAC relation~\cite{Jansen:1996ck}. Here we use the
fermionic boundary states of the Schr\"odinger
functional\footnote{Fermionic observables are constructed
  independently of the number of dynamical flavours $N_f$.} 
to transform
this operator relation to an identity (up to ${\rm O}(a)$) for
fermionic correlation functions which can be
computed on the lattice. This gives a time dependent mass
$m(x_0)$. The mass $m$ is then defined by
\begin{equation}
m=\left\{
\begin{array}{ll}
  m({\scriptstyle\frac{T}{2}}) & \text{$T$ even},\\
  {\scriptstyle\frac12}\left(m({\scriptstyle\frac{T-1}{2}})+m({\scriptstyle\frac{T+1}{2}})\right) 
  & \text{$T$ odd.}
\end{array}
\right.
\end{equation}

To define the step scaling function $\sigma(s,u)$ let $u=\bar{g}^2(L)$
and $m(L)=0$. Then we change the length scale by a factor $s$ and
compute the new coupling $u'=\bar{g}^2(sL)$. The lattice step scaling
function $\Sigma$ at the resolution $L/a$ is defined as
\begin{equation}
  \Sigma(s,u,a/L)=\bar{g}^2(sL)\big\vert_{\bar{g}^2(L)=u,\ m(L)=0}.
\end{equation}
Note that the two conditions fix the two bare parameters $g_0$ and
$\kappa$. The continuum limit $\sigma(s,u)$ can be found by an
extrapolation in $a/L$. That means the computational strategy is as
follows: 
\begin{itemize}
\item[1.] Choose a lattice with $L/a$ points in each direction.
\item[2.] Tune the bare parameters $g_0$ and $\kappa$ such that the
  renormalized coupling $\bar{g}^2(L)$ has the value $u$ and $m(L)=0$.
\item[3.] At the same value of $g_0$ and $\kappa$ simulate a lattice
  with twice the linear size and compute $u'=\bar{g}^2(2L)$. This gives
  $\Sigma(2,u,\frac{a}{L})$.
\item[4.] Repeat steps 1.-3. with different resolutions $L/a$ and
  extrapolate $\frac{a}{L}\rightarrow 0$, which yields $\sigma(2,u)$.  
\end{itemize}

%

%
%
%
\section{Results}

We have performed Monte Carlo simulations on APE100/Quadrics parallel
computers with SIMD architecture and single precision
arithmetics. The size of the machines ranged from 8 to 256 nodes. The
gauge fields and the bermion fields have been generated by hybrid
overrelaxation including microcanonical reflection steps. While we
measured the gauge observables after each update of the fields the
fermionic correlation functions were determined only every 100th to
1000th iteration, since their measurement involves the inversion of
the Dirac operator. The statistical errors have been determined by a
direct computation of the autocorrelation matrix. The largest run has
been performed for the lattice size $L=24$ which took about 12 days on
the largest machine.
\begin{figure}[htb]
\vspace{-20mm}
\begin{center}
\epsfig{file=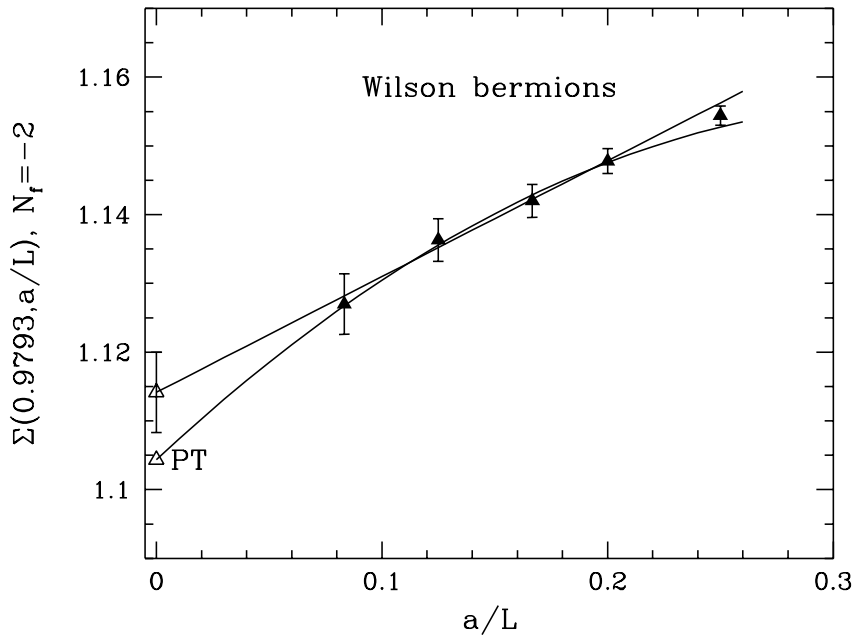,width=12.0cm}
\parbox{12.0cm}{
\vspace{-0.9cm}
\caption{\label{fig3} \sl  Results for the lattice step scaling
  function for $N_f=-2$ at $u=0.9793$ confronted with renormalized
      perturbation theory.}
}
\end{center}
\end{figure}
\begin{figure}[htb]
\vspace{-12mm}
\begin{center}
\epsfig{file=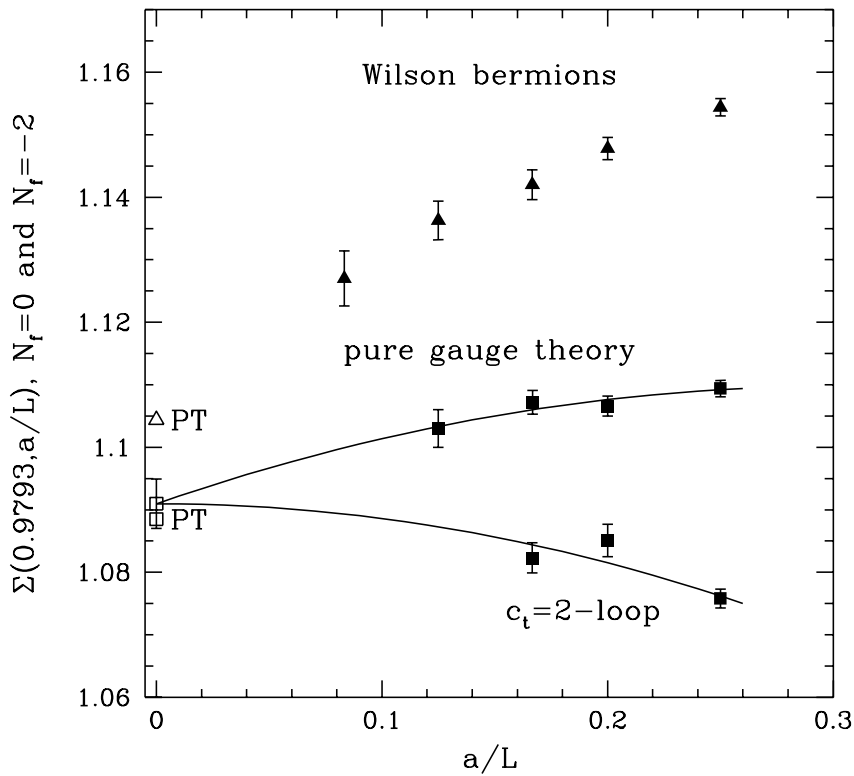,width=12.0cm}
\parbox{12.0cm}{
\vspace{-0.9cm}
\caption{\label{fig4} \sl  Results for the lattice step scaling
  functions for $N_f=0$ in the unimproved and the improved case at
  $u=0.9793$ confronted with renormalized perturbation theory.}
}
\end{center}
\end{figure}

In figure~\ref{fig3} our results for the bermion lattice step scaling
function \\
$\Sigma(2,0.9793,a/L)$  are shown for resolutions 4, 5, 6, 8 and 12
and compared with renormalized perturbation theory. The lattice
artifacts are consistent with ${\rm O}(a)$ effects. Included is a
linear fit to the data which extrapolates to a continuum limit
which lies above the perturbative estimate but is compatible with it. 
Furthermore a linear plus quadratic fit with the constraint that the
continuum limit is given by the perturbative estimate is shown. All
the data points are compatible with this fit. Note that $L/a=4$ has
been ignored in these fits. 

In figure~\ref{fig4} we compare these data with results from the
quenched approximation with the standard Wilson action and with a
perturbatively ${\rm O}(a)$ improved action. Again the most naive
extrapolation of the unimproved data would extrapolate to a value
slightly above the perturbative estimate. However, if we 
use universality, i.e. the agreement of the continuum limit of the two
data sets as a constraint, their joint continuum limit is fully
compatible with perturbation theory. 

Furthermore we observe that in the bermion theory the cut off effects
are larger than in the quenched approximation. 

Since the extrapolation to the continuum limit in the quenched
approximation is much easier in the improved case we plan to study
the ${\rm O}(a)$ improved bermion model. This will also allow us to
compare with the fermionic theory on equal footing.

This work is part of the ALPHA collaboration research program. We
thank DESY for allocating computer time to this project and the DFG
under GK 271 for financial support.
\clearpage

\end{document}